\documentclass{pasj00}
\draft
\SetRunningHead{H.J. Haubold A.M. Mathai R.K. Saxena}{Analysis of Solar Neutrino Data from SuperKamiokande I and II}

\usepackage{times}

\begin{document}

\title{Analysis of Solar Neutrino Data from SuperKamiokande I and II:\\ Back to the Solar Neutrino Problem
\thanks{Dedicated to the memory of Professor Dr. Masatoshi Kitamura, National Astronomical Observatory of Japan, who outstandingly supported annual United Nations workshop for the benefit of the worldwide development of astronomy and astrophysics over a period of time of a quarter of a century
(W. Wamsteker et al. 2004, Developing Basic Space Science World-Wide: A Decade of UN/ESA Workshops, (Dordrecht Boston London: Kluwer Academic Publishers); B.J. Thompson et al. 2009, Putting the "I" in IHY: The United Nations Report for the International Heliophysical Year 2007, (Wien New York: Springer); http://www.iaucomm46.org/content/united-nations-basic-space-science-initiative-unbssi-1991-2012)}}

\author{H.J. \textsc{Haubold}}
\affil{%
Office for Outer Space Affairs, United Nations, Vienna International Centre,\\
P.O. Box 500, A-1400 Vienna, Austria,\\
and Centre for Mathematical Sciences, Pala Campus, Arunapuram P.O., Pala, Kerala-686574, India}
\email{hans.haubold@unvienna.org}

\author{A.M. \textsc{Mathai}}
\affil{Department of Mathematics and Statistics, McGill University,\\
Montreal, Canada H3A 2K6,\\
and Centre for Mathematical Sciences, Pala Campus, Arunapuram P.O., Pala, Kerala-686574, India}
\email{mathai@math.mcgill.ca}

\author{R.K. \textsc{Saxena}}
\affil{Department of Mathematics and Statistics, Jai Narain Vyas University,\\
Jodhpur-342005, India}
\email{ram.saxena@yahoo.com}

\KeyWords{methods: Sun: neutrinos${}$ --- Sun: SuperKamiokande${}$ --- methods: numerical${}$}

\maketitle

\begin{abstract}
We are going back to the roots of the original solar neutrino problem: analysis of data from solar neutrino experiments. The application of standard deviation analysis (SDA) and diffusion entropy analysis (DEA) to the SuperKamiokande I and II data reveals that they represent a non-Gaussian signal. The Hurst exponent is different from the scaling exponent of the probability density function and both Hurst exponent and scaling exponent of the probability density function of the SuperKamiokande data deviate considerably from the value of 0.5 which indicates that the statistics of the underlying phenomenon is anomalous. To develop a road to the possible interpretation of this finding we utilize Mathai's pathway model and consider fractional reaction and fractional diffusion as possible explanations of the non-Gaussian content of the SuperKamiokande data.
\end{abstract}

\section{Introduction}

This paper summarizes briefly a research programme, implemented through UN/ESA/NASA/JAXA workshops since 1991, comprised of five elements (i) standard deviation analysis and diffusion entropy analysis of solar neutrino data 
(\cite{Haubold2012a};
\cite{Haubold2012b}),
(ii) Mathai's entropic pathway model
(\cite{Mathai2005};
\cite{Mathai2007}), 
(iii) fractional reaction and extended thermonuclear functions 
(\cite{Haubold2000};
\cite{Kumar2010}), 
(iv) fractional reaction and diffusion 
(\cite{Saxena2004};
\cite{Saxena2010}), 
and (v) fractional reaction-diffusion 
(\cite{Saxena2007a},
\cite{Saxena2007b},
\cite{Saxena2007c};
\cite{Haubold2011}). 
Boltzmann translated Clausius' second law of thermodynamics "The entropy of the Universe tends to a maximum" into a crucial quantity that links equilibrium and non-equilibrium (time dependent) properties of physical systems and related entropy to probability, S = k log W, which later Einstein called Boltzmann's principle 
(\cite{Brush1976}). 
Based on this principle of physics, Planck found the correct formula for black-body radiation that lead him to the discovery of the elementary quantum of action that initiated the development of quantum theory. Extremizing the Boltzmann entropic functional under appropriate constraints produces the exponential functional form of the distribution for the respective physical quantity. Today a question under intense discussion in statistical mechanics is on how to generalize Boltzmann's entropic functional, if extremized under appropriate constraints, to accommodate power law distribution functions observed so frequently in nature. One of such generalizations is Tsallis statistics (\cite{Tsallis2009}) 
that contains Boltzmann statistics as a special case. Tsallis statistics is characterized by q-distributions which seem to occur in many situations of scientific interest and have significant consequences for the understanding of natural phenomena. One of such phenomena concerns the neutrino flux emanating from the gravitationally stabilized solar fusion reactor 
(\cite{DeglInnocenti1998};
\cite{Wolff2009}). 
R. Davis Jr. established the solar neutrino problem which was resolved by the discovery of neutrino oscillations 
(\cite{Smirnov2003};
\cite{Pulido2010}). 
A remaining question to date is still the quest for more information hidden in solar neutrino records of numerous past and currently operating solar neutrino experiments
(\cite{Oser2012}). 
Greatly stimulated by the question, raised long time ago, by R.H. Dicke "Is there a chronometer hidden deep in the Sun?" 
(\cite{Dicke1978};
\cite{Perry1990}), 
Mathai's research programme on the analysis of the neutrino emission of the gravitationally stabilized solar fusion reactor, focused on non-locality (long-range correlations), non-Markovian effects (memory), non-Gaussian processes (L\'{e}vy), and non-Fickian diffusion (scaling), possibly evident in the solar neutrino records, taking also into account results of helio-seismology and helio-neutrinospectroscopy
(\cite{Goupil2011}). 
The original research programme, devised by Mathai, is contained in three research monographs 
(\cite{Mathai1975}; 
\cite{Mathai1977}; 
\cite{Mathai1978}) 
and the results of this research programme were summarized recently in 
\cite{Mathai2010}.

\section{Solar Neutrino Data}

Over the past 40 years, radio-chemical and real-time solar neutrino experiments have proven to be sensitive tools to test both astrophysical and elementary particle physics models and principles. Solar neutrino detectors (radio-chemical: Homestake, GALLEX + GNO, SAGE, real-time: SuperKamiokande, SNO, Borexino) 
(Oser 2012; Haxton et al. 2012) 
have demonstrated that the Sun is powered by thermonuclear fusion reactions. Two distinct processes, the pp-chain and the sub-dominant CNO-cycle, are producing solar neutrinos with different energy spectra and fluxes (see Figure 1). To date only fluxes from the pp-chain have been measured: $^7Be$, $^8B$, and, indirectly, $pp$. Experiments with solar neutrinos and reactor anti-neutrinos (KamLAND; see 
\cite{Haxton2012}) 
have confirmed that solar neutrinos undergo flavor oscillations (Mikheyev-Smirnov-Wolfenstein (MSW) model; see 
\cite{Pulido2010}). 
Results from solar neutrino experiments are consistent with the Mikheyev-Smirnov-Wolfenstein Large Mixing Angle (MSW-LMA) model, which predicts a transition from vacuum-dominated to matter-enhanced oscillations, resulting in an energy dependent electron neutrino survival probability. Non-standard neutrino interaction models derived such neutrino survival probability curves that deviate significantly from MSW-LMA, particularly in the 1-4 $MeV$ transition region. The mono-energetic 1.44 $MeV$ $pep$ neutrinos, which belong to the pp-chain and whose Standard Solar Model (SSM) predicted flux has one of the smallest uncertainties due to the solar luminosity constraint, are an ideal probe to test these competing non-standard neutrino interaction models in the future 
(\cite{Ludhova2012}).

\begin{figure}
\begin{center}
\FigureFile(80mm,80mm){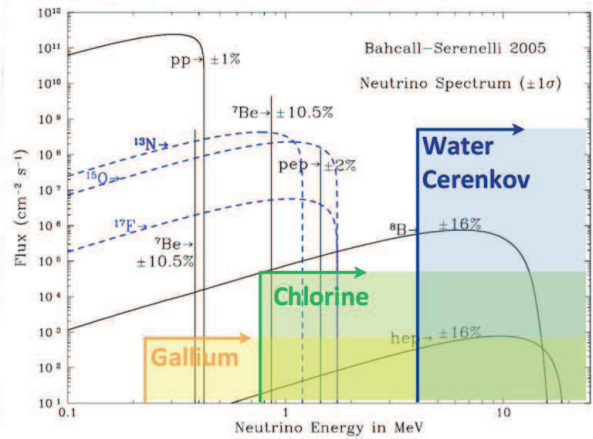}
\end{center}
\caption{The solar neutrino spectrum for the pp-chain and the CNO-cycle and parts of the spectrum that are being detectable by the experiments based on gallium, chlorine, and Cherenkov radiation 
(\cite{Haxton2012}).
}
\end{figure}

\begin{figure}
\begin{center}
\FigureFile(80mm,80mm){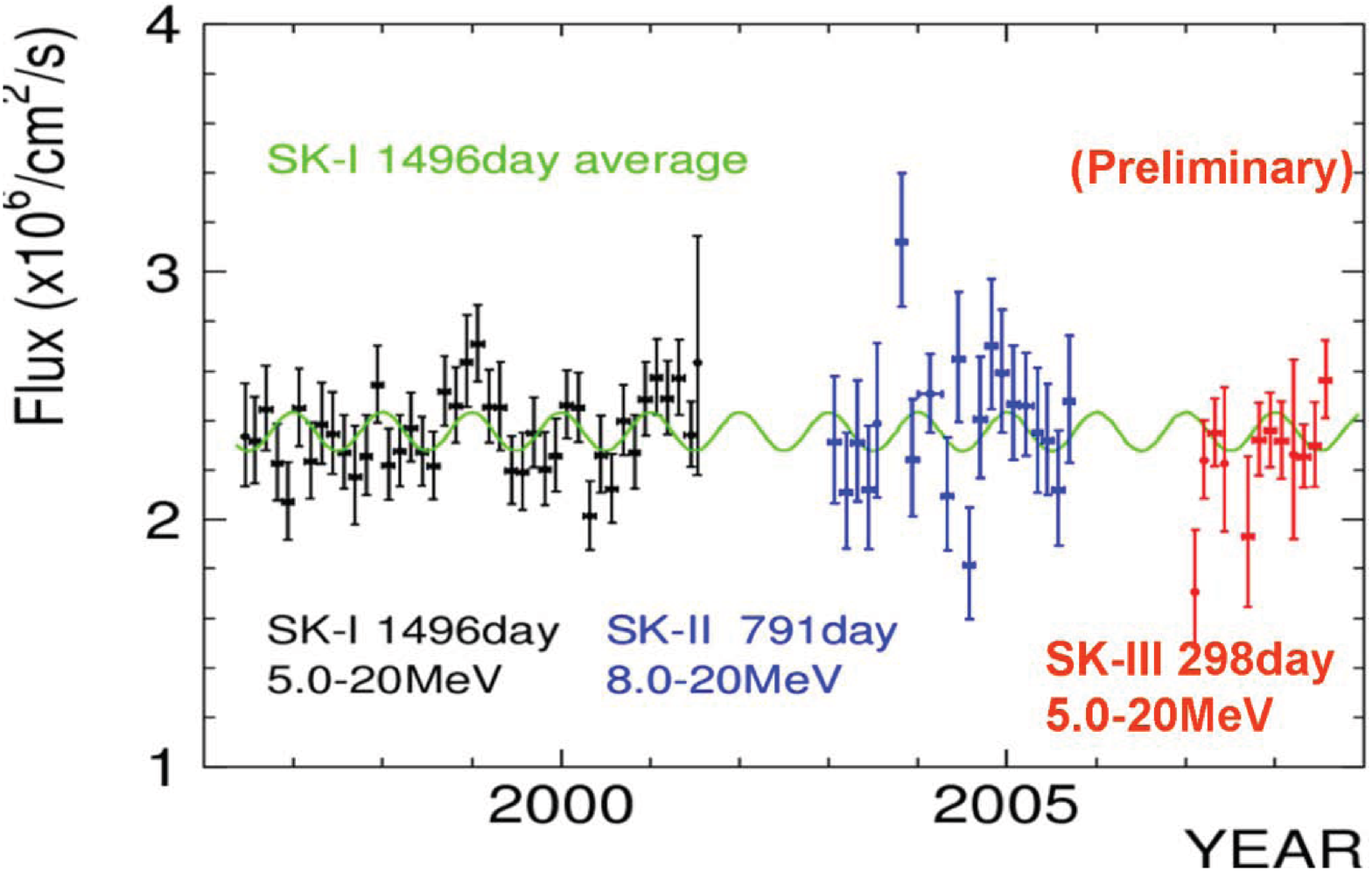}
\end{center}
\caption{The variation of the solar neutrino flux over time as shown in the SuperKamiokande I, II, and III experiments 
(\cite{Yoo2003};
\cite{Cravens2008}; 
\cite{Abe2011}
).}
\end{figure}

\section{Standard Deviation Analysis and Diffusion Entropy Analysis}

For all radio-chemical and real-time solar neutrino experiments, periodic variation in the detected solar neutrino fluxes have been reported, based mainly on Fourier and wavelet analysis methods (standard deviation analysis) 
(\cite{Davis1987};
\cite{Sakurai2008};
\cite{Vecchio2009};
\cite{Vecchio2010}). 
Other attempts to analyze the same data sets, particularly undertaken by the experimental collaborations of real-time solar neutrino experiments themselves, have failed to find evidence for such variations of the solar neutrino flux over time. Periodicities in the solar neutrino fluxes, if confirmed, could provide evidence for new solar, nuclear, or neutrino physics beyond the commonly accepted physics of vacuum-dominated and matter-enhanced oscillations of massive neutrinos (MSW model) that is, after 40 years of solar neutrino experiment and theory, considered to be the ultimate solution to the solar neutrino problem. Specifically, subsequent to the analysis made by the SuperKamiokande collaboration 
(\cite{Yoo2003}; 
\cite{Cravens2008}; 
\cite{Abe2011}), 
the SNO experiment collaboration has painstakingly searched for evidence of time variability at periods ranging from 10 $years$ down to 10 $minutes$. SNO has found no indications for any time variability of the $^8B$ flux at any timescale, including in the frequency window in which $g$-mode oscillations of the solar core might be expected to occur 
(\cite{Aharmim2010}). 
Despite large efforts to utilize helio-seismology and helio-neutrinospectroscopy, at present time there is no conclusive evidence in terms of physics for time variability of the solar neutrino fluxes from any solar neutrino experiment 
(\cite{Pulido2010}; 
\cite{Goupil2011}). 
If such a variability over time would be discovered, for example in the Borexino experiment, a mechanism for a chronometer for solar variability could be proposed based on relations between properties of thermonuclear fusion and g-modes. All above findings encouraged the conclusion that Fourier and wavelet analysis, which are based upon the analysis of the variance of the respective time series (standard deviation analysis: SDA) 
(\cite{Haubold1990}; 
\cite{Haubold1998}), 
should be complemented by the utilization of diffusion entropy analysis (DEA), which measures the scaling of the probability density function (pdf) of the diffusion process generated by the time series thought of as the physical source of fluctuations 
(\cite{Scafetta2002}; 
\cite{Scafetta2010}). 
For this analysis, we have used the publicly available data of SuperKamiokande-I and SuperKamiokande-II (see Figure 2). Such an analysis does not reveal periodic variations of the solar neutrino fluxes but shows how the pdf scaling exponent departs in the non-Gaussian case from the Hurst exponent. Figures 3 to 6 show the Hurst exponents (SDA) and scaling exponents (DEA) for the SuperKamiokande I and II data. SuperKamiokande is sensitive mostly to neutrinos from the $^8B$ branch of the $pp$ nuclear fusion chain in solar burning. Above approximately 4 $MeV$ the detector can pick-out the scattering of solar neutrinos off atomic electrons which produces Cherenkov radiation in the detector. The $^8B$ and rarer $hep$ neutrinos have a spectrum which ends near 20 $MeV$ (see Figure 1).

Assuming that the solar neutrino signal is governed by a probability density function with scaling given by the asymptotic time evolution of a pdf of $x$, obeying the property 

\begin{equation}
p(x,t)=\frac{1}{t^\delta}F(\frac{x}{t^\delta}),
\end{equation}

where $\delta$ denotes the scaling exponent of the pdf. In the variance based methods, scaling is studied by direct evaluation of the time behavior of the variance of the diffusion process. If the variance scales, one would have 

\begin{equation}
\sigma_x^2(t)\sim t^{2H},
\end{equation}

where $H$ is the Hurst exponent. To evaluate the Shannon entropy of the diffusion process at time $t$, 
Scafetta et al. 
(2002; Scafetta 2010) 
defined $S(t)$ as 

\begin{equation}
S(t)=-\int^{+\infty}_{-\infty} dx\;p(x,t) \ln\;p(x,t)
\end{equation}

and with the previous $p(x,t)$ one has

\begin{equation}
S(t)=A+\delta \ln(t),\;\;A=-\int^{+\infty}_{-\infty}dyF(y)\ln F(y)
\end{equation}

The scaling exponent $\delta$ is the slope of the entropy against the logarithmic time scale. The slope is visible in Figures 4 and 6 for the SuperKamiokande data measured for $^8B$ and $hep$. The Hurst exponents (SDA) are $H=0.66$ and $H=0.36$ for $^8B$ and $hep$, respectively, shown in Figures 3 and 5. The $pdf$ scaling exponents (DEA) are $\delta=0.88$ and $\delta=0.80$ for $^8B$ and $hep$, respectively, as shown in Figures 4 and 6. The values for both SDA and DEA indicate a deviation from Gaussian behavior which would require that $H=\delta=0.5$. A preliminary analysis, for SuperKamiokande I data exclusively, was undertaken recently by 
\citet{Haubold2012a}. 
A test computation for the application of SDA and DEA to data that are known to exhibit non-Gaussian behavior have been published recently by 
\citet{Haubold2012b}. 
In this test, SDA and DEA, applied to the magnetic field strength fluctuations recorded by the Voyager-I spacecraft in the heliosphere clearly revealed the scaling behavior of such fluctuations as previously already discovered by non-extensive statistical mechanics considerations that lead to the determination of the non-extensivity q-triplet 
(\cite{Tsallis2009}).

\begin{figure}
\begin{center}
\FigureFile(80mm,80mm){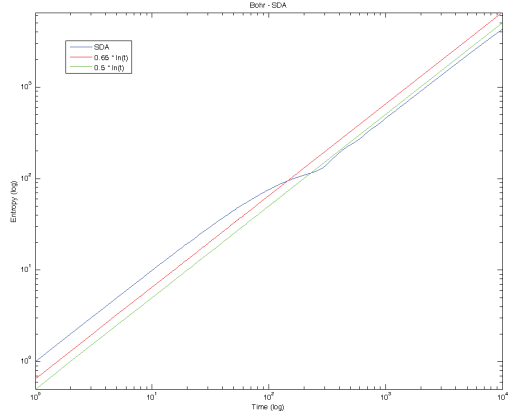}
\end{center}
\caption{The Standard Deviation Analysis (SDA) of the $^8B$ solar neutrino data from the SuperKamiokande I and II experiment.}
\end{figure}

\begin{figure}
\begin{center}
\FigureFile(80mm,80mm){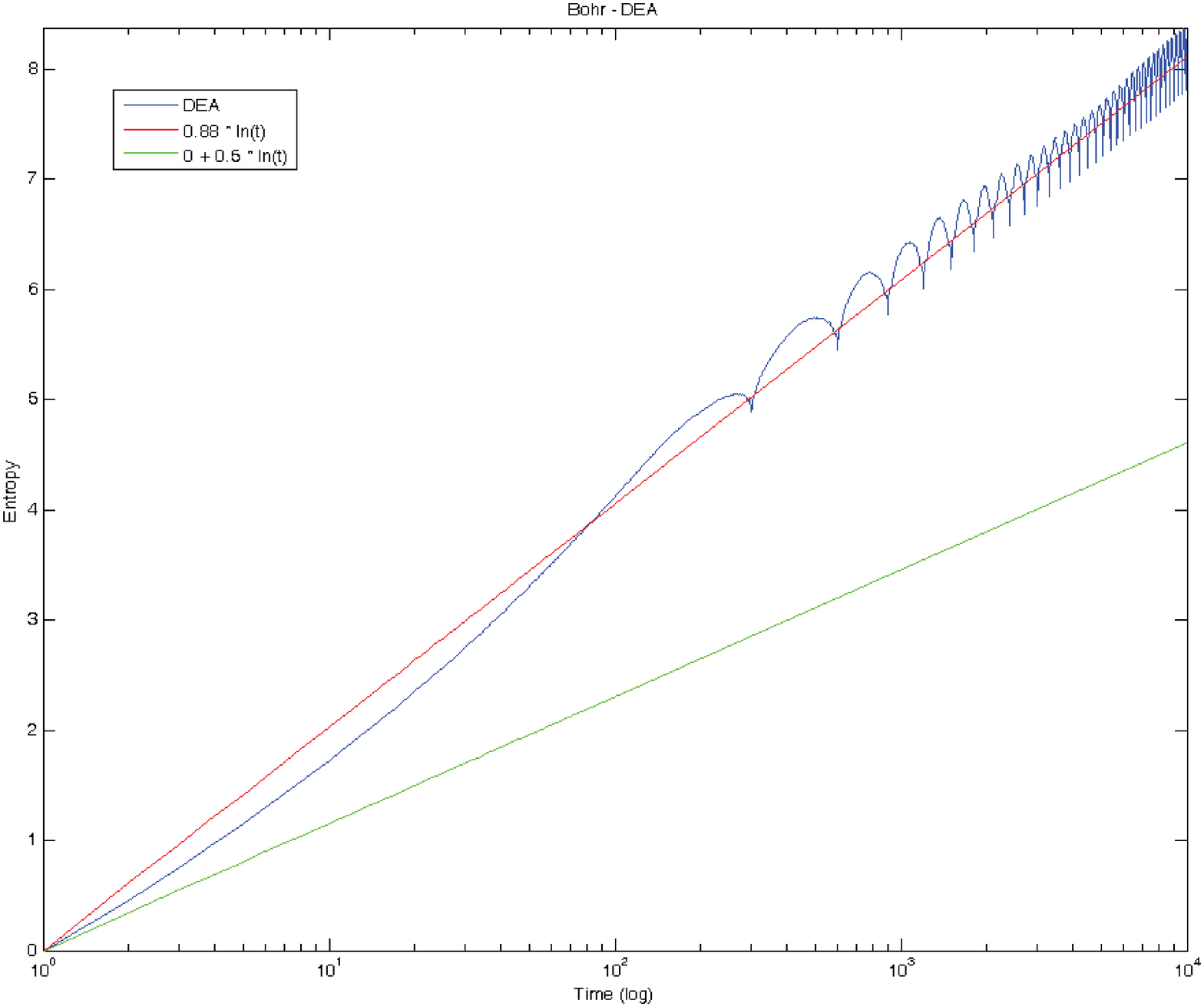}
\end{center}
\caption{The Diffusion Entropy Analysis (DEA) of the $^8B$ solar neutrino data from the SuperKamiokande I and II experiment.}
\end{figure}

\begin{figure}
\begin{center}
\FigureFile(80mm,80mm){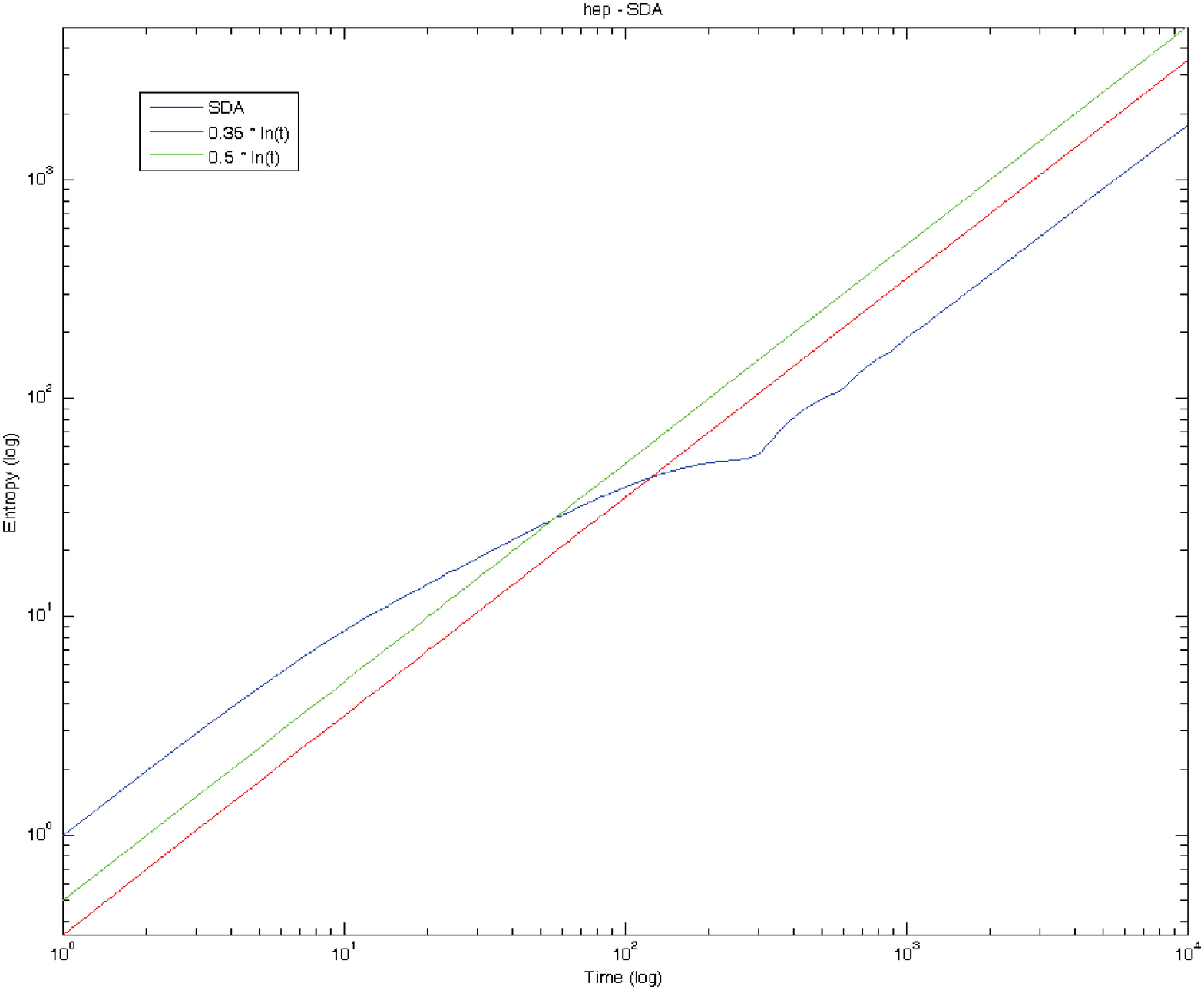}
\end{center}
\caption{The Standard Deviation Analysis (SDA) of the $hep$ solar neutrino data from the SuperKamiokande I and II experiment.}
\end{figure}

\begin{figure}
\begin{center}
\FigureFile(80mm,80mm){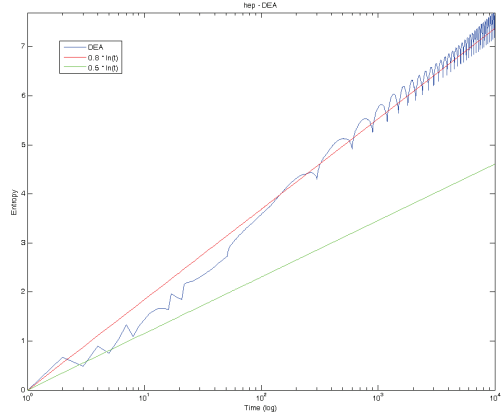}
\end{center}
\caption{The Diffusion Entropy Analysis (DEA) of the $hep$ solar neutrino data from the SuperKamiokande I and II experiment.}
\end{figure}

\section{Mathai's Entropic Pathway Model}

From a general point of view of fitting experimental data to mathematical functions, a model, which moves from the generalized type-1 beta family to the type-2 beta family to the generalized gamma family to the generalized Mittag-Leffler family and eventually to the L\'{e}vy distributions, has been developed by Mathai  
\cite{Mathai2005}. 
All these different parametric families of functions are connected through Mathai's pathway parameter $\alpha>1$. To generalize Shannon's entropy to an entropic pathway, Mathai introduced the generalized entropy of order $\alpha$ that is also associated with Shannon (Boltzmann-Gibbs), Renyi, Tsallis, and Harvrda-Charvat entropies 
(\cite{Mathai2007};
\cite{Mathai2010}). 
Applying the maximum entropy principle with normalization and energy constraints to Mathai's entropic functional, the corresponding parametric families of distributions of generalized type-1 beta, type-2 beta, generalized gamma, generalized Mittag-Leffler, and L\'{e}vy are obtained in the following form

\begin{equation}
M_2(f)=\frac{\int^{+\infty}_{-\infty}dx[f(x)]^{2-\alpha} -1}{\alpha - 1}\;\;\alpha\neq 1, \alpha < 2,
\end{equation}

\begin{equation}
f(x)=c_1[1-\beta(1-\alpha)x^\delta]^{1/(1-\alpha)}
\end{equation}

with $\alpha<1$ for type-1 beta, $\alpha>1$ for type-2 beta, $\alpha \rightarrow 1$ for gamma, and $\delta=1$ for Tsallis statistics. In principle, any entropic functional in Mathai's pathway can be tested through the above Diffusion Entropy Analysis against experimental data. The deviation of the statistical properties of the SuperKamiokande data analyzed above from Gaussian should be captured by equation (6). 

\section{Fractional Reaction and Extended Thermonuclear Functions}

Solar nuclear reactions, producing neutrinos, occur preferably between nuclei in the high-energy tail of the energy distribution and are sensitive to deviations from the standard equilibrium thermal energy distribution (Maxwell-Boltzmann distribution) 
(\cite{Critchfield1972};
\cite{DeglInnocenti1998};
\cite{Kumar2010}). 
Reaction and relaxation processes in thermonuclear plasmas are governed by ordinary differential equations of the type 

\begin{equation}
\frac{dN(t)}{dt}=c\;N(t)
\end{equation}

for exponential behavior. The quantity $c$ is a thermonuclear function which is governed by the average of the Gamow penetration factor over the Maxwell-Boltzmannian velocity distribution of reacting species and has been extended to incorporate more general distributions than the normal distribution 
(\cite{Mathai2010}). 
The coefficient $c$ itself can be considered to be a statistical quantity subject to accommodating a distribution of its own 
(\cite{Mathai2007}). 
To address non-exponential properties of a reaction or relaxation process, the first-order time derivative can be replaced formally by a derivative of fractional order in the following way 
(\cite{Mathai2010})

\begin{equation}
N(t)=N_0-c^\nu\;_0D_t^{-\nu}\;N(t),
\end{equation}

where $_0D_t^{-\nu}$ denotes a Riemann-Liouville fractional integral operator, and the solution can be represented in terms of Mittag-Leffler functions $E_\nu$ by 

\begin{equation}
N(t)=N_0E_\nu(-c^\nu t^\nu).
\end{equation}

Considering $c$ to be a random variable itself, $N(t)$ is to be taken as $N(t\mid c)$ and can be written as 

\begin{equation}
N(t\mid c)=N_0t^{\mu-1}E_{\nu,\mu}^{\gamma+1}(-c^\nu t^\nu),\;\;\mu>0, \gamma >0, \nu>0,
\end{equation}

which represents a generalized Mittag-Leffler function, and is a random variable having a gamma type density 

\begin{equation}
g(c)=\frac{\omega^\mu}{\Gamma(\mu)}\;c^{\mu-1}e^{-\omega c}\;\;\omega>0, 0<c<\infty, \mu>0,
\end{equation}

with $\mu / \omega$ is the mean value of $c$. The integration of $N(t \mid  c)$ over $g(c)$ gives the unconditional density,  as

\begin{equation}
N(t)=\frac{N_0}{\Gamma(\mu)}t^{\mu-1}[1+b(\alpha-1)t^\nu]^{-1/(\alpha-1)},
\end{equation}

with $\gamma +1=1/(\alpha-1), \alpha>1 \rightarrow  \gamma=(\alpha-2)/(\alpha-1)$
and $\omega^{-\nu}=b(\alpha-1), b>0,$ which corresponds to Tsallis statistics for $\mu=1, \nu=1, b=1,$ and $\alpha=q>1$, physically meaning that the common exponential behavior is replaced by a power-law behavior, including L\'{e}vy statistics. Both the translation of the standard reaction equation (7) to a fractional reaction equation (8) and the probabilistic interpretation of such equations lead to deviations from the exponential behavior to power law behavior expressed in terms of Mittag-Leffler functions (9) or, as can be shown for equation (12), to power law behavior in terms of H-functions 
(\cite{Mathai2010}). 
H-functions are representable in terms of Mellin-Barnes integrals of the product of gamma functions and are therefore suited to represent statistics of products and quotients of independent random variables thus providing a very useful tool in presenting a new perspective on the statistics of random variables 
(\cite{Cottone2010}). 

\section{Fractional Diffusion and the Joint Action of Reaction and Diffusion}

In recent time, an analytic approach to non-conventional reaction and diffusive transport by taking into account fractional space and time derivatives has been developed 
(\cite{DelCastillo2008}). 
The probability density function for the above SuperKamiokande data is non-Gaussian and exhibits stretched power-law tails as can be shown by further exploring equations (6), (9), and (12). In order to model these analytic findings, a transport model for the pdf, based on fractional diffusion that includes both non-local and non-Gaussian features was proposed 
(\cite{Mathai2010}). 
Reaction and diffusion in the solar thermonuclear fusion plasma are non-linear phenomena that may be subject to non-Fickian transport (non-locality), non-Markovian effects (memory), and non-Gaussian scaling (L\'{e}vy). Fractional diffusion operators are integro-differential operators that incorporate the former three phenomena in a natural way and may be, in this regard, constitute spatio-temporal elements of fundamental theory of physics. This issue is currently under intense research. Continues time random walk (CTRW) balance equations (master equations) with temporal memory, generation/destruction terms, and spatio transport/relaxation elements yield non-linear fractional reaction-diffusion equations whose solutions are a focus of current research and only very special cases have been dealt with so far. Equally difficult to reveal is the interplay between fractional reaction and fractional diffusion in such non-linear equations. This difficulty is amplified by the fact that various definitions of fractional operators exist (Riemann-Liouville, Caputo, Weyl, Gruenwald-Letnikov, Riesz-Feller, ...). At this point of time there is no general understanding under which specific mathematical and physical conditions a probabilistic interpretation can be given to unified fractional reaction-diffusion equations. And this difficulty is even further amplified by the observation that the replacement of integer order with fractional order time derivatives changes the fundamental concept of time and violates the principle that time evolution (change) is time translation and that fractional order space derivatives are bridging the respective differential equation between the case of diffusion equation and wave equation.

\section{Conclusion}

The use of solar neutrino detection records, by analyzing the average neutrino flux of the experiments, have lead to the discovery of new elementary particle physics, the MSW effect, and thus resolved the solar neutrino problem. This confirmed that the Standard Solar Model is implementing physical principles correctly. The quest for the variation of the solar neutrino flux over time remains an open question. Additionally, the utilization of standard deviation analysis (scaling of the variance) and diffusion entropy analysis (scaling of the pdf) lead to the discovery of an unknown phenomenon related to non-equilibrium signature in the gravitationally stabilized solar fusion reactor as explored by looking at Mathai's pathway model and taking into account fractional reaction and fractional diffusion and possibly a combination of both of them.

\medskip
The authors acknowledge the cooperation of Dr. Alexander Haubold of the Department of Computer Science, Columbia University, New York. The authors would like to thank the Department of Science and Technology, Government of India, New Delhi, for the financial assistance under project SR/S4/MS:287/05.

{}

\begin{thebibliography}{}
\bibitem [Abe et al.\ (2011)]{Abe2011} Abe, K. et al. 2011, Solar neutrino results in Super-Kamiokande-III, Physical Review D 83, 052010\\
\bibitem [Aharmim et al.\ (2010)] {Aharmim2010} Aharmim, B. et al. 2010, Searches for high-frequency variations in the $^8B$ solar neutrino flux at the Sudbury Neutrino Observatory, The Astrophysical Journal 710, 540\\
\bibitem [Brush\ (1976)]{Brush1976} Brush, S.G. 1976, Irreversibility and indeterminism: Fourier to Heisenberg, Journal of the History of Ideas 37, 603\\
\bibitem [Cravens et al.\ (2008)]{Cravens2008} Cravens, J.P. et al. 2008, Solar neutrino measurements in Super-Kamiokande-II, Physical review D 78, 032002\\
\bibitem[Cottone et al.\ (2010)]{Cottone2010} Cottone, G., Di Paola, $\&$ Metzler, R. 2010, Fractional calculus approach to the statistical characterization of random variables and vectors, Physica A 389, 909\\
\bibitem[Critchfield\ (1972)]{Critchfield1972} Critchfield, Ch.L. 1972, Analytic forms of the thermonuclear function, in Cosmology, Fusion and Other Matters, F. Reines (Ed.), Colorado Associated University Press, 186\\
\bibitem[Davis et al.\ (1987)]{Davis1987} Davis Jr., R., Cleveland, B.T., $\&$ Rowley, J.K. 1987, Variations in the solar neutrino flux, BNL 39602, 1\\
\bibitem [Degl`Innocenti et al.\ (1998)]{DeglInnocenti1998} Degl`Innocenti, S., Fiorentini, G., Lissia, M., Quarati, P., $\&$ Ricci, B. 1998, Helioseismology can test the Maxwell-Boltzmann distribution, Physics Letters B 441, 291\\
\bibitem [Del-Castillo\ (2008)]{DelCastillo2008} Del-Castillo-Negrete, D. 2008, Fractional diffusion models of anomalous transport, in Anomalous Transport: Foundations and Applications, R. Klages, G. Radons, and I.M. Sokolov (Eds.), Wiley-VCH, Weinheim, 163\\
\bibitem [Dicke\ (1978)]{Dicke1978} Dicke, R.H. 1978, Is there a chronometer hidden deep in the Sun?, Nature 276, 676\\
\bibitem [Goupil\ (2011)]{Goupil2011} Goupil, M.J., Lebreton Y., Marques, J.P., Samadi, R., $\&$ Baudin, F. 2011, Open issues in probing interiors of solar-like oscillating main sequence stars: 1. From the Sun to nearly suns, Journal of Physics: Conference Series 271, 012031\\
\bibitem [Haubold et al.\ (2012a)]{Haubold2012a} Haubold, A., Haubold, H.J., $\&$ Kumar, D. 2012a, Solar neutrino records: Gauss or non-Gauss is the question, arXiv: 1202.1549v1 [physics.gen-ph]\\
\bibitem [Haubold et al.\ (2012b)]{Haubold2012b} Haubold, A., Haubold, H.J., $\&$ Kumar, D. 2012b, Heliosheath: Diffusion entropy analysis and nonextensivity q-triplet, arXiv: 1202.3417v1 [physics.gen-ph]\\
\bibitem [Haubold \& Gerth (1990)]{Haubold1990} Haubold, H.J., $\&$ Gerth, E. 1990, On the Fourier spectrum analysis of the solar neutrino capture rate, Solar Physics 127, 347\\
\bibitem[Haubold \& Mathai(1998)]{Haubold1998} Haubold, H.J., $\&$ Mathai, A.M. 1998, Wavelet analysis of the new solar neutrino capture rate for the Homestake experiment, Astrophysics and Space Science 258, 201\\
\bibitem[Haubold \& Mathai (2000)]{Haubold2000} Haubold, H.J., $\&$ Mathai, A.M. 2000, The fractional kinetic equation and thermonuclear functions, Astrophysics and Space Science 273, 53\\
\bibitem[Haubold \& Mathai (2011)]{Haubold2011} Haubold, H.J., Mathai, A.M., $\&$ Saxena, R.K. 2011, Further solutions of fractional reaction-diffusion equations in terms of the H-function, Journal of Computational and Applied Mathematics 235, 1311\\
\bibitem[Haxton et al.\ (2012)]{Haxton2012} Haxton, W.C., Hamish Robertson, R.G., $\&$ Serenelli, A.M. 2012, Solar neutrinos: Status and prospects, arXiv: 1208.5723v1 [astro-ph.SR]\\
\bibitem[Kumar \& Haubold (2010)]{Kumar2010} Kumar, D., $\&$ Haubold, H.J. 2010, On extended thermonuclear functions through the pathway model, Advances in Space Research 45, 698\\
\bibitem[Ludhova et al.\ (2012)]{Ludhova2012} Ludhova, L. et al. 2012, Solar neutrino physics with Borexino I, arXiv: 1205.2989v1 [hep-ex]\\
\bibitem[Mathai\ (2005)]{Mathai2005} Mathai, A.M. 2005, A pathway to matrix-variate gamma and normal densities, Linear Algebra and its Applications 396, 317\\
\bibitem[Mathai \& Haubold (2007)]{Mathai2007} Mathai, A.M., $\&$ Haubold, H.J. 2007, Pathway model, superstatistics, Tsallis statistics, and a generalized measure of entropy, Physica A 375, 110\\
\bibitem[Mathai \& Pederzoli (1977)]{Mathai1977} Mathai, A.M. $\&$ Pederzoli, G. 1977, Characterizations of the Normal Probability Law, (New Delhi Bangalore Bombay: Wiley Eastern Limited)\\
\bibitem[Mathai \& Rathie (1975)]{Mathai1975} Mathai, A.M., $\&$ Rathie, P.N. 1975, Basic Concepts in Information Theory and Statistics: Axiomatic Foundations and Applications, (New York London Sydney Toronto: John Wiley $\&$ Sons)\\
\bibitem[Mathai \& Saxena (1978)]{Mathai1978} Mathai, A.M., $\&$ Saxena, R.K. 1978, The H-function with Applications in Statistics and Other Disciplines, (New York London Sydney Toronto: John Wiley $\&$ Sons)\\
\bibitem[Mathai et al.\ (2010)]{Mathai2010} Matahi, A.M., Saxena, R.K., $\&$ Haubold, H.J. 2010, The H-Function: Theory and Applications, (New York Dordrecht Heidelberg London: Springer)\\  
\bibitem[Oser\ (2012)]{Oser2012} Oser, S.M. 2012, An experimentalist's overview of solar neutrinos, Journal of Physics: Conference Series 337, 012056\\
\bibitem[Perry\ (1990)]{Perry1990} Perry, Ch.A. 1990, Speculations on a solar chronometer for climate, NASA Conference Publication 3086, 357\\
\bibitem[Pulido et al.\ (2010)]{Pulido2010} Pulido, J., Das, C.R., $\&$ Picariello, M. 2010, Remaining inconsistencies with solar neutrinos: Can spin flavor precession provide a clue?, Journal of Physics: Conference Series 203, 012086\\
\bibitem[Sakurai et al.\ (2008)]{Sakurai2008} Sakurai, K., Haubold, H.J., $\&$ Shirai, T. 2008, The variation of the solar neutrino fluxes over time in the Homestake, GALLEX (GNO) and the Super-Kamiokande Experiments, Space Radiation 5, 207\\
\bibitem [Saxena et al.\ (2004)]{Saxena2004} Saxena, R.K., Mathai, A.M., $\&$ Haubold, H.J. 2004, Unified fractional kinetic equation and a fractional diffusion equation, Astrophysics and Space Science 209, 299\\
\bibitem[Saxena et al.\ (2007a)]{Saxena2007a} Saxena, R.K., Mathai, A.M., $\&$ Haubold, H.J. 2007a, Fractional reaction-diffusion equations, Astrophysics and Space Science 305, 289\\
\bibitem[Saxena et al.\ (2007b)]{Saxena2007b} Saxena, R.K., Mathai, A.M., $\&$ Haubold, H.J. 2007b, Reaction-diffusion systems and nonlinear waves, Astrophysics and Space Science 305, 297\\
\bibitem[Saxena et al.\ (2007c)]{Saxena2007c} Saxena, R.K., Mathai, A.M., $\&$ Haubold, H.J. 2007c, Solution of generalized fractional reaction-diffusion equations, Astrophysics and Space Science 305, 305\\
\bibitem[Saxena et al.\ (2010)]{Saxena2010} Saxena, R.K., Mathai, A.M., $\&$ Haubold, H.J. 2010, Solutions of certain fractional kinetic equations and a fractional diffusion equation, Journal of Mathematical Physics 51, 103506\\
\bibitem[Scafetta\ (2010)]{Scafetta2010} Scafetta, N. 2010, Fractal and Diffusion Entropy Analysis of Time Series: Theory, concepts, applications and computer codes for studying fractal noises and L\'{e}vy walk signals, (Saarbruecken: VDM Verlag Dr. Mueller)\\
\bibitem[Scafetta \& Latora (2002)]{Scafetta2002} Scafetta, N., Latora, V., $\&$ Grigolini, P. 2002, Levy statistics in coding and non-coding nucleotide sequences, Physics Letters A 299, 565\\
\bibitem[Smirnov\ (2003)]{Smirnov2003} Smirnov, A.Yu. 2003, The MSW effect and solar neutrinos, arXiv: 0305106 [hep-ph]\\
\bibitem[Tsallis\ (2009)]{Tsallis2009} Tsallis, C. 2009, Introduction to Nonextensive Statistical Mechanics: Approaaching a Complex World, (New York: Springer)\\
\bibitem[Vecchio \& Carbone (2009)]{Vecchio2009} Vecchio, A., $\&$ Carbone, V. 2009, Spatio-temporal analysis of solar activity: Main periodicities and period length variations, Astronomy and Astrophysics 502, 981\\
\bibitem[Vecchio et al.\ (2010)]{Vecchio2010} Vecchio, A., Laurenza, M., Carbone, V., $\&$ Storini, M. 2010, Quasi-biennial modulation of solar neutrino flux and solar and galactic cosmic rays by solar cyclic activity, The Astrophysical Journal Letters 709, L1\\
\bibitem[Wolff\ (2009)]{Wolff2009} Wolff, Ch.L. 2009, Effects of a deep mixed shell on solar $g$-modes, $p$-modes, and neutrino flux, The Astrophysical Journal 701, 686\\
\bibitem[Yoo\ (2003)]{Yoo2003} Yoo J. et al. 2003, Search for periodic modulations of the solar neutrino flux in Super-Kamiokande-I, Physical Review D 68, 092002\\

\end{thebibliography}
\end{document}